\documentclass[draft,prb,preprint,showpacs]{revtex4}

\usepackage{amsmath,amsfonts,amssymb,bm}
\usepackage{dcolumn}
\usepackage[final]{graphicx}
\usepackage{bm}

\begin{document}

\title{Optical near-field mapping of excitons and biexcitons in naturally occurring semiconductor quantum dots}

\author{Ulrich Hohenester}\email{ulrich.hohenester@uni-graz.at}
\affiliation{Institut f\"ur Theoretische Physik,
  Karl--Franzens--Universit\"at Graz, Universit\"atsplatz 5,
  8010 Graz, Austria}

\author{Guido Goldoni}
\author{Elisa Molinari}
\affiliation{INFM--$S^3$ and
Dipartimento di Fisica, Universit\`a di Modena e Reggio Emilia,
Via Campi 213/A, 41100 Modena, Italy
}

\date{March 8, 2004}

\begin{abstract}
We calculate the near-field optical spectra of excitons and
biexcitons in semiconductor quantum dots naturally occurring at
interface fluctuations in GaAs-based quantum wells, using a
non-local description of the response function to a spatially
modulated electro-magnetic field. The relative intensity of the
lowest, far-field forbidden excitonic states is predicted; the
spatial extension of the ground biexciton state is found in
agreement with recently published experiments.
\end{abstract}


\maketitle


Optical manipulation of states in quantum dots (QDs) is a key
towards the implementation of quantum information processing in
semiconductors.~\cite{barenco:95,troiani:00,biolatti:00}
Important features of QDs are not
only that excitations are fully discrete in energy resulting in
long coherence times,~\cite{bonadeo:98,zrenner:00} but also that they are governed by
the Coulomb interaction between electrons and holes which gives
rise to specific few-particle aggregates as a result of
photoexcitation. Based on this unique mechanism, all-optical
conditional control of exciton and biexciton states were proposed
and demonstrated for quantum computing schemes in QD structures.~\cite{troiani:00,li:03}

The use of local probes for a detailed understanding of these
interacting states and their real-space distribution is thus of
great interest for their engineering, and might open the way to
the important goal of direct optical manipulation of individual
states in coupled structures. The so-called natural QDs
---or terraces---, where quantum confinement is induced by local
monolayer fluctuations in the thickness of a semiconductor quantum
well, have sofar been excellent laboratories for the study of
these novel phenomena because quantization is combined with a
large oscillator strength associated to the large active volume of
the QD.~\cite{bastard.89} The huge progress in probe preparation has lead to
resolutions which in this class of systems have reached the scale
of the constituent quantum states.~\cite{guest:01,matsuda:02a,%
matsuda:02b,matsuda:03}

In this letter, we investigate theoretically the optical response
of excitons and biexcitons in natural QDs in near-field
experiments. We calculate the spatial maps of the exciton and
biexciton low-energy states, taking into account the non-local
susceptibility of the electromagnetic (EM) field, and show that the spatial extension
depends on the correlated state, being narrower for biexcitons, in
agreement with experimental findings. We also calculate the
oscillator strengths of the lowest, far-field forbidden excitonic
states, and show that they can be comparable to the ground state,
thus allowing for their observation at the presently achievable
resolution scale.


{\em Exciton and biexciton states.}\/ For excitonic complexes
bound to interface fluctuations, where the confinement
considerably exceeds the excitonic Bohr radius, center-of-mass and
relative motion of the electron-hole pairs can be decoupled. Then,
the in-plane part of the exciton and biexciton wavefunctions, $\Psi^x$ and
$\bar\Psi^b$, respectively, can be approximately written
as~\cite{zimmermann:97}

\begin{eqnarray}
\Psi^x(\bm r_1,\bm r_a) &\cong& \phi_o(\bm r_1,\bm r_a)\,\Phi^x(\bm R_x)\nonumber\\
\bar\Psi^b(\bm r_1,\bm r_2,\bm r_a,\bm r_b) &\cong&
\bar\phi_o(\bm r_1,\bm r_2,\bm r_a,\bm r_b)\,\bar\Phi^b(\bm R_b)\,.
\end{eqnarray}

\noindent Here, $\phi_o$ and $\bar\phi_o$ are the exciton and
biexciton wavefunctions for an ideal quantum well, whereas
$\Phi^x$ and $\bar\Phi^b$ are the envelope functions accounting
for interface fluctuations. We denote electrons with $\bm r_1$ and
$\bm r_2$, holes with $\bm r_a$ and $\bm r_b$, and finally the
exciton and biexciton center-of-mass coordinates with $\bm R_x$
and $\bm R_b$. The specific forms of $\phi_o$ and $\bar\phi_o$ are
taken from Kleinman~\cite{kleinman:83} for a 5 nm thick quantum well,
i.e., the usual
two-dimensional exciton state $\propto\exp-kr_{1a}$, $k=1.94$ in
units of the Bohr radius $a_o$, and the biexciton state
$\psi(r_{1a},r_{1b},r_{2a},r_{2b})\chi(r_{ab})$, which consists of
trial wavefunctions $\psi$ and $\chi$ associated to the attractive
electron-hole~\cite{hylleraas:47} and repulsive
hole-hole~\cite{brinkman:73} interactions, respectively; $r_{ij}$
is the distance between particles $i$ and $j$.

We consider a prototypical interface-fluctuation confinement of
rectangular shape with dimensions $100\times 70$
nm$^2$, and monolayer fluctuations of a 5 nm thick
well.~\cite{matsuda:02b,matsuda:03} The exciton confinement is then given 
by the well-width dependent electron and hole single-particle energies 
along $z$ convoluted with the probabilities $p_i(\bm r-\bm R)$ of finding an
electron or hole at distance $\bm r$ from the center-of-mass
coordinate $\bm R$,~\cite{zimmermann:97} with a straightforward
generalization to biexcitons ($\bm\tau$ and $\bar{\bm\tau}$ denote the
phase space for electrons and biexcitons, respectively)

\begin{eqnarray}
p_i(\bm r-\bm R)&=&\int d\bm\tau\,\delta(\bm R-\bm R_x)\delta(\bm r-\bm r_i)\,
\phi_o^2(\bm r_1,\bm r_a)\nonumber\\
\bar p_i(\bm r-\bm R)&=&\int d\bar{\bm\tau}\,
\delta(\bm R-\bm R_b)\delta(\bm r-\bm r_i)\,
\bar\phi_o^2(\bm r_1,\bm r_2,\bm r_a,\bm r_b)\,.\nonumber\\
\end{eqnarray}

Figure 1 shows the effective confinement potential for excitons
(solid line) and biexcitons (dashed line). The insets show that
for excitons the electron probability distribution $p_e(\bm r)$
extends over the effective Bohr radius $k^{-1}a_o$ whereas
$p_h(\bm r)$ is strongly peaked around $\bm 0$; in contrast, for
the biexcitons the weaker Coulomb binding results in a strong
delocalization of the two-exciton complex.~\cite{kleinman:83}
As consequence, the biexciton center-of-mass motion is confined
within a significantly smaller region.

The resulting two-dimensional Schr\"odinger equation for the
exciton and biexciton center-of-mass wavefunctions $\Phi^x(\bm
R_x)$ and $\bar\Phi^b(\bm R_b)$, respectively, are solved through a
real-space discretization and numerical diagonalization on a grid
of typical dimensions $64\times 64$, similarly to the procedure
described in Refs.~\onlinecite{pistone:03}. The computed exciton and
biexciton states exhibit symmetries reminiscent of the two-dimensional
box-like confinement,~\cite{comment.energies} i.e., an $s$-like
exciton groundstate (fig.~2a), two $p$-like excited states of lowest energy
with nodes along $x$ (fig.~2b) and $y$ (not shown), and two nodes along
$x$ for the third excited exciton state (fig.~2c); finally, the biexciton groundstate
(fig.~2d) indeed shows  a much stronger localization than the exciton
groundstate (fig.~2a).

{\em Near-field spectra.}\/ When the EM-field is modulated on the
scale of the relevant quantum states, the non-local response of
the system must be taken into account.~\cite{mauritz} In addition
to the relaxation of the far-field selection rules due to the
different symmetry of the field with respect to the quantum
states, spatial coherence may give rise to interference effects,
so that collected spatial maps may be non-trivially related to the
localization of the excitonic wavefunctions. We compute the
near-field spectra analogously to the procedure described in
Refs.~\onlinecite{mauritz,simserides:00} for an EM-field
distribution $\xi(\bm R_{\rm tip}-\bm r)$ of Gaussian shape
centered around the tip position $\bm R_{\rm tip}$. For the
excitons the local absorption spectra at a given exciton energy
$E_x$ are then given by the square modulus of the convolution of
$\xi$ with the exciton wavefunction
$\Phi^x$.~\cite{simserides:00,pistone:03} For the biexciton we
have to be more specific of how the system is excited. We shall
assume that the QD is initially populated by the ground state
exciton and that the near-field tip probes the transition to the
biexciton states. This situation approximately corresponds to that
of Ref.~\onlinecite{matsuda:02b,matsuda:03} with non-resonant excitation 
in the non-linear power regime. The local spectra are then
proportional to~\cite{unpublished} $\int d\bm r\; \xi(\bm R_{\rm
tip}-\bm r)\langle x|\hat\psi_h(\bm r)\hat\psi_e(\bm r)|b\rangle$,
with $x$ and $b$ denoting the exciton and biexciton states of
eq.~(1), respectively, and $\hat\psi_{e(h)}(\bm r)$ is the usual
fermionic field operator for electrons (holes). Then, the square
modulus of

\begin{equation}\label{eq:mu}
\int d{\bm R}\,\Phi^x(\bm R)\,\mu(\bm r-\bm R)\,\bar\Phi^b(\frac{\bm r+\bm R}2)
\end{equation}

\noindent convoluted with $\xi$ gives the optical near-field
spectra corresponding to the transition from $x$ to $b$, with
$\mu(\bm r-\bm R)=\int d\bm\tau\,\delta(\bm R-\bm R_x)\,\phi_o(\bm
r_1,\bm r_a)\bar\phi_o(\bm r,\bm r_1,\bm r,\bm r_a)$ giving the
probability of exciting an electron-hole pair at $\bm r$ when an
exciton is located at $\bm R$ (see fig.~1e)~\cite{unpublished}.


In the second and third rows of fig.~2 we report our calculated
optical near-field spectra for spatial resolutions of 25 and 50
nm. It should be noted that the first (fig.~2b) and second excited
state (not shown) are dipole forbidden, but have large oscillator
strengths for both resolutions. Note also that, as a result of
interference effects, the spatial maps at finite spatial
resolutions differ somewhat from the wavefunction maps,
particularly for the excited states: the apparent localization is
weaker and, in (c), the central lobe is very weak for both
resolutions. Finally, we observe that for the smaller spatial
resolution the biexciton ground state depicts a stronger degree of
localization than the exciton one, in nice agreement with the
recent experiment of Matsuda et al.~\cite{matsuda:02b,matsuda:03}.

In order to be more quantitative and show if near-field
experiments may distinguish the dipole forbidden transitions, we
have calculated the total absorbed power (i.e., the incoherently
summed intensity of the maps exemplified in fig.~2) as a function
of the spatial resolution. Figure 3 shows that the intensity of
the lowest dipole-forbidden state is a substantial fraction of the
ground state intensity, and larger than the next dipole allowed
excited states up to resolutions comparable with the QD linear
dimensions. Indeed, preliminary results indicate that $p$-like
structures have been seen in the sample of
Ref.~\onlinecite{matsuda:02a}.~\cite{private}

We thank K. Matsuda for helpful discussions.
This work has been supported in part by the EU under the TMR
network ``Exciting'', the Austrian science fund FWF under project
P15752--N08, the Italian Minister for University and Research
under project FIRB {\em Quantum phases of ultra-low electron
density semiconductor heterostructures},\/ and by INFM I.T. Calcolo
Parallelo (2003).

\newpage

\begin{figure}
\caption{Confinement potential along $x$ for the center-of-mass
motion of excitons (solid line) and biexcitons
(dashed line).~\cite{comment.energies}
The insets report the exciton probability distributions $p_i(\bm r)$
for (a) electrons and (b) holes, the biexciton probability
distributions $\bar p_i(\bm r)$ for (c) electrons and (d) holes,
and (e) $\mu(\bm r)$ the probability of creating a second
electron-hole pair at distance $\bm r$ from the exciton center of
mass. The reduced probability at the center of (d) is attributed
to the repulsive part $\chi$ of the trial
wavefunction.~\cite{kleinman:83} }
\end{figure}

\begin{figure*}
\caption{(a--d) Real-space map of the square modulus of the
wavefunctions for  the exciton (a) ground state, (b) first and (c)
third excited  state, and (d) the biexciton ground
state;~\cite{comment.energies} the
dashed lines indicate the boundaries of the assumed interface
fluctuation. (a'--d') Near-field spectra for a spatial resolution
of 25 nm and (a''--d'') 50 nm, as computed according to
Refs.~\onlinecite{mauritz,simserides:00} and eq.~\eqref{eq:mu}.
The FWHM of the assumed EM-field distribution is indicated in the
2$^{\mbox{\scriptsize nd}}$ and 3$^{\mbox{\scriptsize rd}}$ row}
\end{figure*}

\begin{figure}
\caption{Total absorbed power of the first (black, dashed line),
second (gray, dashed line), and third (black, solid line) excited
exciton state as a function of the spatial resolution of $\xi$.
For all resolutions the spectra are normalized to the ground state
absorption. The first and second excited excitons are far-field
forbidden, whereas the oscillator strengths of the third excited state is
approximately one ninth of that of the ground state.}
\end{figure}

\end{document}